# Mode conversion and orbital angular momentum transfer among multiple modes by helical gratings

Liang Fang, Jian Wang*
*Corresponding author: jwang@hust.edu.cn

*Abstract*—The coupling feature of helical gratings (HGs) has been studied on flexible conversion between two orbital angular momentum (OAM) modes by both the transverse and longitudinal modulation of HGs. Apart from one-to-one OAM exchange, HGs can achieve OAM transfer among multiple modes, provided both the helix and phase matching conditions are well satisfied for successive coupling. It is contributed from the fold number of modulation fringes of HGs and the difference multiplication of propagation constants between two of the successive modes. Based on the coupled mode theory, we investigate the mode conversion and OAM transfer among three modes using a 3-fold HGs in a ring-core fiber, and numerically simulate it by transmission matrix. Our work about the multiple modes transferring can be regarded as an extension of mode coupling for optical gratings, and might flourish the mode conversion, as well as OAM manipulation in optics fields.

## I. INTRODUCTION

Optical gratings have been developed into a mature technology and been broadly applied to optical communications and fiber sensing in recent decades [1-5]. It is based on mode coupling due to refractive index modulation that makes it possible to achieve free conversion between any two modes supported in fibers. Apart from reflecting fundamental core modes, coupling from the fundamental core mode to abundant cladding modes has been successfully achieved by long-period gratings (LPGs), fiber Bragg gratings (FBGs), and tilted LPGs or FBGs in common fibers [6-8], which highly depends on wavelength due to the waveguide dispersion. In recent years, with the development of space-division multiplexing (SDM), mode-division multiplexing (MDM), and orbital angular momentum (OAM) multiplexing as new techniques to expand the ever-increasing transmission capacity [9-12], fiber gratings fabricated in few-mode fibers (FMFs) or multimode fibers (MMFs) have been used to gain mode conversion between the fundamental mode and higher-order modes as mode couplers or multiplexer/de-multiplexer [13-15]. Especially, for OAM mode coupling and conversion, microbend gratings with a metallic block with grooves are adopted in vortex fiber and FMF [16, 17], but they are not all-fiber systems due to the bulky structure. In previous work, we have presented a theoretical study on flexible generation, conversion, and exchange of fiber-guided OAM modes using helical gratings (HGs) with index modulation formed in ring-core fibers (RCFs), which can be regarded as an all-fiber method of manipulating OAM modes [18]. However, almost all previous reports are about one-to-one mode coupling as well as OAM exchange. In this paper, we present a new coupling feature of mode conversion and OAM transfer among multiple modes by HGs. One-to-two mode coupling ever proposed requires superposing two sets of gratings in references[19, 20], but seemingly just the phase matching condition is highlighted. It is still vague that what the explicit condition of transverse modulation needs to be met and how it affects the coupling for high order mode with high efficiency. Here by utilizing an uniform 3-fold HGs, we theoretically investigate the mode conversion and OAM transfer among three modes in RCF with detail, and numerically simulate it based on the method of transmission matrix.

It has been well known that vortex fiber or RCF can support multiple OAM modes with the lowest radial order. These OAM modes have large difference of effective refractive index ($\Delta n_{eff} > 10^{-4}$) among the vector mode groups that may tend to degenerate into linearly polarized (LP) mode in conventional fibers. Therefore, it is possible to transmit several OAM modes with low inter-mode crosstalk in these fibers [21-23]. Based on previous work [18], our investigation on OAM transfer among multiple modes is still restricted in RCF. It can be contributed from the fold number of modulation of HGs and the difference multiplication of propagation constants between two of the successive OAM modes supported in RCF. Our work can be regarded as an extension of mode coupling for optical gratings, and might flourish the mode conversion, as well as OAM manipulation in optics fields. One can believe that the phenomenon of mode coupling and OAM transfer among multiple modes by fiber gratings could become universal in large dimensional fibers such as MMFs and RCFs that support several or lots of modes, because there are some of them that we can find to meet both the helix and phase matching conditions. As for fabricating helical gratings (HGs), it could be done through rotating the fiber when writing them by point-by-point method or a phase mask in about one micrometer range [24-26], as well as by ion implantation [27], and thermo-diffusion in nitrogen-doped silica-core fibers [28].

## II. Coupling Coefficient

Firstly, we focus on the discussion of coupling coefficient. In high-contrast-index fibers, the expression of coupling coefficient cannot been directly formulized associated with two independent OAM modes in reference [18]. Before giving its exact expression, we discuss the constituent of fiber-guided OAM modes, especially in the high-contrast-index fibers, such as the hollow RCF. Due to the effect of spin-orbit interaction [29, 30], OAM modes cannot be simply described by a single matrix. Instead, it is a superposition of the major OAM mode and a minor one, of which topological charge numbers has a difference of $\pm 2$. The amplitude ratio of them is $|(1+\eta)/(1-\eta)|$, where $\eta$ is the proportion between azimuthal and radial field components of the corresponding fiber vector modes [29]. We can denote these OAM modes as follows,

$$|+1,+j\rangle \to \xi\left[(1+\eta)|+1,+j\rangle + (1-\eta)|-1,+(j+2)\rangle\right], \quad \text{(1-a)}$$
$$|-1,-j\rangle \to \xi\left[(1+\eta)|-1,-j\rangle + (1-\eta)|+1,-(j+2)\rangle\right], \quad \text{(1-b)}$$
$$|-1,+j\rangle \to \xi\left[(1+\eta)|-1,+j\rangle + (1-\eta)|+1,+(j-2)\rangle\right], \quad \text{(1-c)}$$
$$|+1,-j\rangle \to \xi\left[(1+\eta)|+1,-j\rangle + (1-\eta)|-1,-(j-2)\rangle\right], \quad \text{(1-d)}$$

where $\xi = 1/\sqrt{2(1+\eta^2)}$ is normalized coefficient. Generally, the value of $|1-\eta|$ approaches 0 in low-contrast-index fiber, but increases with waveguides' index contrast degree. Note that in the text below, for convenience, we still use the denotation $|s_n, l_n\rangle$ to indicate the impure OAM modes. It has been known that in transmissive HGs, mode coupling only occurs between two OAM modes with the same polarization [18]. The coupling coefficient of any two OAM modes using HGs modulated with the same function as [18] can be exactly written as,

$$\kappa_{nm} = \frac{\omega \varepsilon_0 n_2^2 \Delta n}{8 a_2^2} (1+\eta_n \eta_m) \cdot$$
$$\int_{a_1}^{a_2} \int_0^{2\pi} F_n(r)^* \cdot F_m(r) \exp\left[j(l_m - l_n + l)\phi\right] r^3 dr d\phi, \quad (2)$$

where $*$ denotes operation of conjugate transpose, $\omega$ and $\varepsilon_0$ are the angular frequency and dielectric constant in vacuum, respectively, and $F_n(r)$ corresponds to the radial function of electric field, $a_1$ and $a_2$ are the inside and outside radii of RCF, respectively, $n_2$ is the refractive index of the ring-core waveguide layer. $\Delta n$ denotes the modulation strength of HGs.

## III. Coupled Mode Equations

Coupled-mode equations of all possible modes involved in mode coupling, of which amplitudes are denoted with $A_1$, $A_2$, $A_3$, ..., $A_k$, respectively, along the $z$ axis can be expressed as,

$$\begin{cases} \dfrac{dA_1}{dz} = j\kappa_{12} A_2 \exp(-j2\delta_{12} z) \\ \sum_{i=3}^{k} \left[ \dfrac{dA_{i-1}}{dz} = j\kappa_{i-1,i-2}^* A_{i-2} \exp(j2\delta_{i-2,i-1} z) \\ \qquad\qquad + j\kappa_{i-1,i} A_i \exp(-j2\delta_{i-1,i} z) \right], \\ \dfrac{dA_k}{dz} = j\kappa_{k,k-1}^* A_{k-1} \exp(j2\delta_{k-1,k} z) \end{cases} \quad (3)$$

where $\kappa_{i-1,i}$ is defined by (2), $\kappa_{i-1,i} = \kappa_{i-1,i}^*$ and

$$\delta_{i-1,i} = \frac{1}{2}(\beta_{i-1} - \beta_i - K_{i-1}), \quad (4)$$

with

$$K_{i-1} = 2\pi\sigma l/t\Lambda, \ (t=1,2,3,...), \ (i=2,3,4,...,k). \quad (5)$$

One can see that mode coupling among $A_1$, $A_2$, $A_3$, ..., $A_k$ has the property of transitivity that is the mechanism of the mode conversion and OAM transfer among multiple modes. For given HGs $\langle \sigma, |l\rangle\rangle$ with grating period $\Lambda$, OAM transfer among modes $|s_{i-1}, l_{i-1}\rangle$ only occurs when the conditions of both phase matching $\beta_{i-1} - \beta_i = K_{i-1}$, and helix matching $l_i - l_{i-1} = l$ are successively satisfied at the same time. It needs to be emphasized that only the transmissive HGs possess the function of multiple-modes synchronous coupling, but the reflective HGs do not, according to the phase matching condition.

This phenomenon of mode coupling and OAM transfer among multiple modes can occur in a 3-fold HGs inscribed in the hollow RCF. The parameters of RCF are designed as follows: the inside and outside radii of RCF are $a_1 = 7.50$ μm, and $a_2 = 9.00$ μm, respectively, and the refractive index of innermost layer and equivalently infinite cladding are set as $n_1 = 1$ (refractive index of air) and $n_3 = 1.62$, that of ring-core waveguide is $n_2 = 1.668$. This RCF can support vector modes (HEnm/EHnm) with high mode order $n$ close to 9 and only the first radial order $m=1$ (1550nm). The wavelength dependence of effective refractive index for all these fiber vector modes including TM$_{01}$ and TE$_{01}$ is shown in Fig. 1. One can see that the vector mode groups that may tend to degenerate into LP modes have been completely split. Therefore, it is easy to find several OAM modes that are combined with even and odd vector modes with a $\pi/2$ phase shift meet both the phase and helix matching conditions for mode coupling and OAM transfer among multiple modes. At 1550nm, we find that three OAM modes $|+1,+1\rangle$, $|+1,+4\rangle$ and $|+1,+7\rangle$ corresponding to combination of odd and even vector modes HE21, HE51 and HE81 with a $\pi/2$ phase shift, respectively, can be selected for realization of this coupling phenomenon. The character of HGs is ruled by $\langle +1,3\rangle$, and thus the grating constants $K_1 = 3\pi/\Lambda$, $(t=2)$ and $K_2 = 6\pi/\Lambda$, $(t=1)$. Fig. 2 illustrates the concept of mode coupling and OAM transfer from mode $|+1,+1\rangle$ to $|+1,+4\rangle$ and then to $|+1,+7\rangle$ with the 3-fold HGs, where OAM modes with different topological charge is described by the helical phase with corresponding fold number.

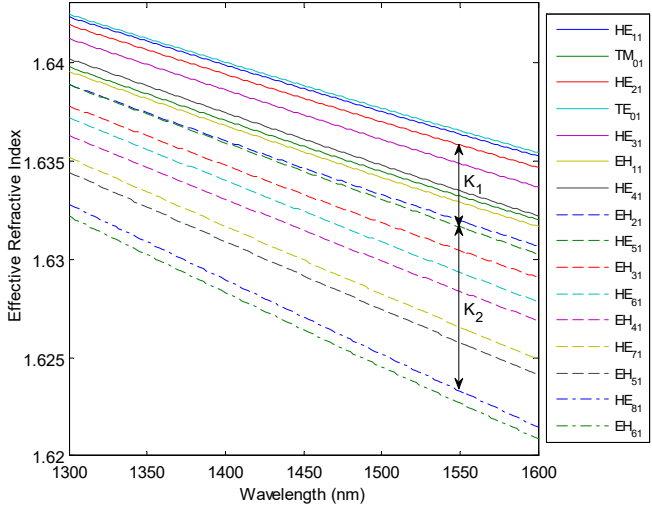

Fig. 1. Wavelength dependence of effective refractive index of vector modes in the designed hollow RCF.

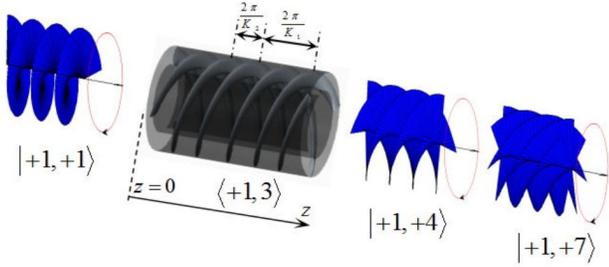

Fig. 2. OAM transfer from modes $|+1,+1\rangle$ to $|+1,+4\rangle$ and then to $|+1,+7\rangle$ with the HGs $\langle+1,3\rangle$.

For the mode coupling and OAM transfer among three modes described by (3) with $k=3$, the detuning factors of mode coupling between $|+1,+1\rangle$ and $|+1,+4\rangle$ are

$$\delta_{12} = \frac{1}{2}(\beta_1 - \beta_2 - K_1), \quad (6)$$

and for $|+1,+4\rangle$ and $|+1,+7\rangle$,

$$\delta_{23} = \frac{1}{2}(\beta_2 - \beta_3 - K_2). \quad (7)$$

The corresponding coupling coefficients with normalization calculated are $\kappa_{12}/\Delta n \simeq \kappa_{23}/\Delta n \simeq 1.05 \times 10^{-3}$ ( $nm^{-1}$ ), which are the largest values since that the helix matching condition is satisfied for the integral express in (2). The helix matching can be equivalently illustrated by the interaction of transverse phase between OAM modes and HGs shown in Fig. 3.

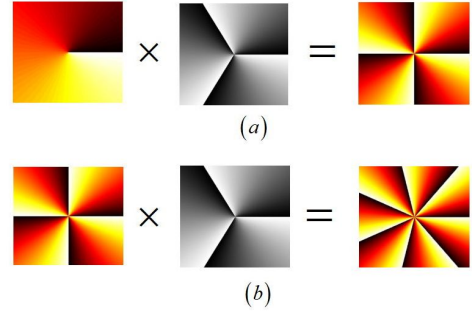

Fig. 3. Interaction of transverse phase between OAM modes and HGs. (a), from $|+1,+1\rangle$ to $|+1,+4\rangle$ through HGs $\langle+1,3\rangle$; (b), from $|+1,+4\rangle$ to $|+1,+7\rangle$ through HGs $\langle+1,3\rangle$. The hot and gray patterns correspond to OAM modes and HGs, respectively.

To solve the coupled-mode equations in (3) when $k=3$, defining new amplitudes of those coupled OAM modes $T_1$, $T_2$ and $T_3$ near resonance region associated with original amplitudes $A_1$, $A_2$ and $A_3$,

$$\begin{aligned} T_1 &= A_1 \exp(-j\sigma_1 z) \\ T_2 &= A_2 \exp(-j\sigma_2 z) \\ T_3 &= A_3 \exp(-j\sigma_3 z) \end{aligned}, \quad (8)$$

where

$$\sigma_1 = \beta_2 - \beta_3 - 2K_1, \quad (9\text{-a})$$
$$\sigma_2 = \beta_1 - \beta_3 - 3K_1, \quad (9\text{-b})$$
$$\sigma_3 = \beta_1 + \beta_2 - 2\beta_3 - 5K_1, \quad (9\text{-c})$$

the coupled-mode equations defined by (3) can be rewritten as,

$$\begin{cases} \dfrac{dT_1}{dz} = -j\sigma_1 T_1 + j\kappa_{12} T_2 \\ \dfrac{dT_2}{dz} = j\kappa_{12} T_1 - j\sigma_2 T_2 + j\kappa_{23} T_3 \\ \dfrac{dT_3}{dz} = j\kappa_{23} T_2 - j\sigma_3 T_3 \end{cases}. \quad (10)$$

The complex coupled-mode equations of multiple modes coupling in (3) cannot be easily solved to obtain an analytic solution, especially for more than 3 modes involved in coupling. Here we use the method of transmission matrix to numerically solve this kind of coupled-mode equations [31]. As for (10), it can be expressed by the form of matrix

$$\frac{d}{dz}\mathbf{T}(z) - \mathbf{M} \cdot \mathbf{T}(z) = 0, \quad (11)$$

where $\mathbf{T}(z) = [T_1(z), T_2(z), T_3(z)]^\mathrm{T}$ with superscript T meaning transpose, and

$$\mathbf{M} = \begin{bmatrix} -j\sigma_1 & j\kappa_{12} & 0 \\ j\kappa_{12} & -j\sigma_2 & j\kappa_{23} \\ 0 & j\kappa_{23} & -j\sigma_3 \end{bmatrix}. \quad (12)$$

Integrating (11) in the grating region from $0$ to $L$, where $L$ is grating length, it is rewritten as

$$\mathbf{T}(L) = \exp(\mathbf{S}) \cdot \mathbf{T}(0). \quad (13)$$

Defining

$$\mathbf{S} = \int_0^L \mathbf{M}\, dz = L \cdot \mathbf{M}. \quad (14)$$

Finally, the output matrix through the HGs can be obtained by

$$\mathbf{T}(L) = \mathbf{V} \exp(\mathbf{D}) \mathbf{V}^{-1} \cdot \mathbf{T}(0), \quad (15)$$

where $\mathbf{D}$ is a diagonal matrix constituted by the eigen values of matrix $\mathbf{S}$, matrix $\mathbf{V}$ is constituted by all the corresponding eigen vectors, and $\mathbf{T}(0)$ is the initialization matrix.

Under complete resonance condition, i.e., $\delta_{12} = \delta_{23} = 0$, the transmission matrix $\mathbf{Q}$ related to the power exchange of central wavelength for the OAM mode coupling among three modes along the transmission direction can be obtained as

$$\mathbf{Q} = \begin{bmatrix} 1 - \dfrac{2\kappa_{12}^2}{\gamma^2}\sin^2\left(\dfrac{1}{2}\gamma z\right) & j\dfrac{\kappa_{12}}{\gamma}\sin(\gamma z) & \dfrac{2\kappa_{12}\kappa_{23}}{\gamma^2}\sin^2\left(\dfrac{1}{2}\gamma z\right) \\ j\dfrac{\kappa_{12}}{\gamma}\sin(\gamma z) & \cos(\gamma z) & j\dfrac{\kappa_{23}}{\gamma}\sin(\gamma z) \\ \dfrac{2\kappa_{12}\kappa_{23}}{\gamma^2}\sin^2\left(\dfrac{1}{2}\gamma z\right) & j\dfrac{\kappa_{23}}{\gamma}\sin(\gamma z) & 1 - \dfrac{2\kappa_{23}^2}{\gamma^2}\sin^2\left(\dfrac{1}{2}\gamma z\right) \end{bmatrix}, \quad (16)$$

where $\gamma^2 = \kappa_{12}^2 + \kappa_{23}^2$, and thus the output power $\mathbf{P}(z) = \mathbf{Q}(\gamma z) \cdot \mathbf{P}(0)$. In Table 1, the proportions are listed in some specific values of $\gamma z$, where $\chi = \kappa_{12}^2 / \kappa_{23}^2$.

TABLE I
POWER PROPORTIONS OF MODE COUPLING AMONG THREE MODES IN CENTRAL RESONANCE WAVELENGTH

| $\mathbf{P}(0)$ | $[1\ 0\ 0]^T$ | $[0\ 1\ 0]^T$ | $[0\ 0\ 1]^T$ |
|---|---|---|---|
| $\gamma z = (2n+1)\pi/2$ $(n=0,1,2...)$ | $1:\chi(1+\chi):\chi$ | $\chi:0:1$ | $\chi:\chi(1+\chi):1$ |
| $\gamma z = n\pi$ $(n=1,2...)$ | $\dfrac{(1-\chi)^2}{(1+\chi)^2}:0:\dfrac{4\chi}{(1+\chi)^2}$ | $0:1:0$ | $\dfrac{4\chi}{(1+\chi)^2}:0:\dfrac{(1-\chi)^2}{(1+\chi)^2}$ |

## IV. SIMULATION RESULTS

As for the three-modes coupling in the designed hollow RCF, based on the coupling conditions of both phase matching and helix matching as well as the coupling efficiency under $\gamma z = \pi/2$ at the resonance wavelength of 1550 nm, the parameters of HGs $\langle +1, 3\rangle$ need to be designed as grating period $\Lambda = 557.9\ \mu m$, grating length $L = 3.0$ cm, and modulation strength $\Delta n = 3.5\times 10^{-5}$, respectively. The power exchange relations among three modes are present in Fig. 4. In Fig. 4(a), the initial input power is $\mathbf{P}(0) = [0\ 1\ 0]^T$, it shows that the power of middle OAM mode $|+1,+4\rangle$ is almost equally coupled into two adjacent OAM modes $|+1,+1\rangle$ and $|+1,+7\rangle$ at 1550 nm because of the nearly equal coupling coefficients $\kappa_{12} \simeq \kappa_{23}$ according to (16) and Table 1. The bandwidth of mode coupling is inversely proportional to the number of grating periods $N$, for coupling from $|+1,+4\rangle$ to $|+1,+1\rangle$, $N_1 = 3L/2\Lambda$, whereas for coupling from $|+1,+4\rangle$ to $|+1,+7\rangle$, $N_2 = 2N_1 = 3L/\Lambda$. Therefore, the conversion bandwidth for the former is twice as much as that for the latter. In Figs. 4(b) and (c), the initial inputs are $\mathbf{P}(0) = [1\ 0\ 0]^T$ and $[0\ 0\ 1]^T$, respectively, one can see that the power flowing between two non-adjacent modes are indirectly achieved through the middle mode $|+1,+4\rangle$. The synchronous coupling among three modes are determined by both the helix and phase matching conditions of multiple modes coupled successively, and the power exchange follows the principle of energy conservation.

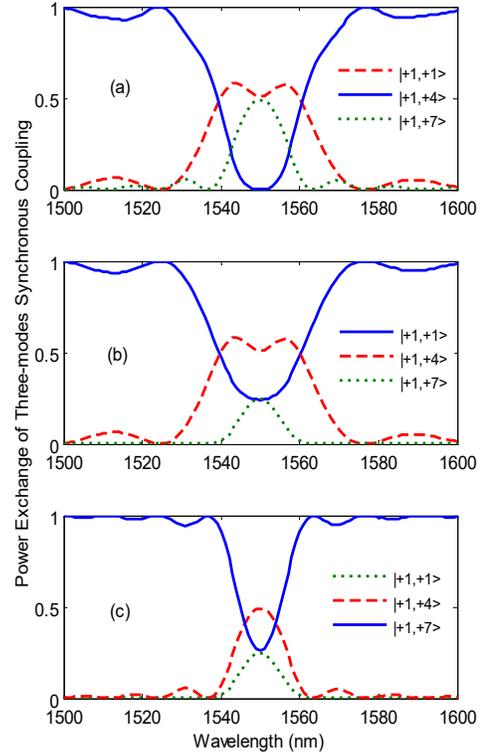

Fig. 4. Power transmission spectra for three-modes coupling with initial inputs of (a) $\mathbf{P}(0) = [0\ 1\ 0]^T$, (b) $\mathbf{P}(0) = [1\ 0\ 0]^T$, and (c) $\mathbf{P}(0) = [0\ 0\ 1]^T$, $\gamma z = \pi/2$.

In Fig. 5, we increase the grating length to $L = 6.0$ cm so that the coupling efficiency is under $\gamma z = \pi$, whereas other parameters and initial inputs remain unchanged. One can see that for $\mathbf{P}(0) = [0\ 1\ 0]^T$, power flowing from $|+1,+4\rangle$ to $|+1,+1\rangle$ and $|+1,+7\rangle$ is over coupled at the resonance wavelength of 1550 nm. Consequently, there are two sidebands for both coupled two modes near the resonance wavelength in Fig. 5(a). It is interesting that for $\mathbf{P}(0) = [1\ 0\ 0]^T$ and $[0\ 0\ 1]^T$ in Figs. 5(b) and (c), respectively, power can completely flow between two non-adjacent modes through the middle mode $|+1,+4\rangle$ at the resonance wavelength, whereas near the resonance wavelength, there are also two sidebands for the middle mode. Additionally, we present the relationship between the power exchange and the grating strength described by $\Delta n \cdot L$ shown in Fig. 6 to explicate the dependence of three-mode coupling on the sensitivity to the grating length and modulation strength when fabricating the HGs. One can see that the power exchange among three OAM modes exhibits a periodic variation with the grating strength. Especially, at the resonant wavelengths, i.e., $\gamma z = n\pi/2$ $(n=1,2...)$, the power proportions of three-mode coupling accords with Table 1. This phenomenon of synchronous coupling among three modes could become universal in FMFs or MMFs where multiple modes are supported, as long as both the helix and phase matching conditions are successively met at the time. The helical modulation of HGs involves into the overlap integral of coupled modes' orthogonal fields in (2) so that makes the coupling coefficient large enough, analogous to adopting the non-uniform modulation in tilted gratings [19], which is the requirement to realize coupling between higher order modes.

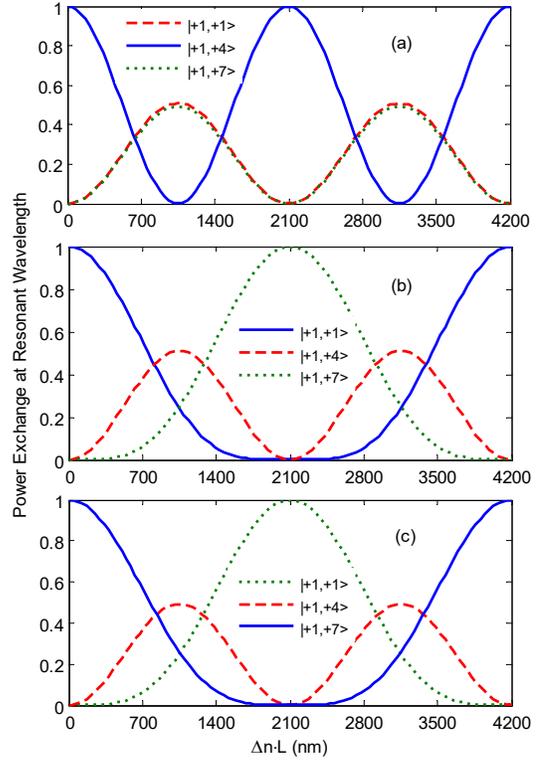

Fig. 6. Power exchange at resonant wavelength versus the values of $\Delta n \cdot L$ with initial inputs of (a) $\mathbf{P}(0) = \begin{bmatrix} 0 & 1 & 0 \end{bmatrix}^T$, (b) $\mathbf{P}(0) = \begin{bmatrix} 1 & 0 & 0 \end{bmatrix}^T$, and (c) $\mathbf{P}(0) = \begin{bmatrix} 0 & 0 & 1 \end{bmatrix}^T$.

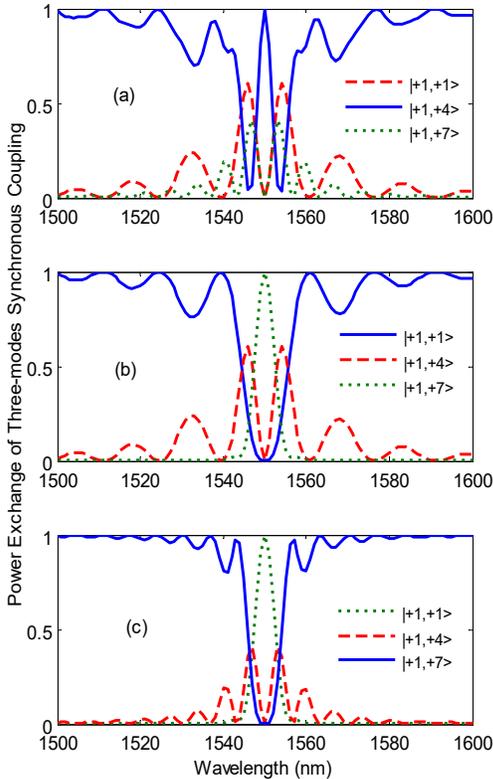

Fig. 5. Power transmission spectra for three-modes coupling with initial inputs of (a) $\mathbf{P}(0) = \begin{bmatrix} 0 & 1 & 0 \end{bmatrix}^T$, (b) $\mathbf{P}(0) = \begin{bmatrix} 1 & 0 & 0 \end{bmatrix}^T$, and (c) $\mathbf{P}(0) = \begin{bmatrix} 0 & 0 & 1 \end{bmatrix}^T$. $\gamma z = \pi$.

## V. CONCLUSION

We present an interesting coupling phenomenon of mode conversion and OAM transfer among multiple modes. As an example, we give detailed theoretical analyses on the three modes coupling in hollow RCF using transmissive HGs. The results show regular power exchanging or flowing and OAM transfer among these OAM modes. It is expected that the multiple modes coupling mechanism might enable flexible mode conversion and OAM manipulation in optics fields.

ACKNOWLEDGMENT

This work was supported by the National Basic Research Program of China (973 Program) under grant 2014CB340004, the National Natural Science Foundation of China (NSFC) under grants 11274131, 11574001 and 61222502, the National Program for Support of Top-notch Young Professionals, the Program for New Century Excellent Talents in University under grant NCET-11-0182, the Wuhan Science and Technology Plan Project under grant 2014070404010201, the seed project of Wuhan National Laboratory for Optoelectronics (WNLO), the open program from State Key Laboratory of Advanced Optical Communication Systems and Networks under grant 2016GZKF0JT007, and the open projects foundation of Yangtze Optical Fiber and Cable Joint Stock Limited Company (YOFC) under grant SKLD1504.